\documentclass[journal]{IEEEtran}
\usepackage{amsmath}
\usepackage{amsfonts}
\usepackage{amssymb}
\usepackage[ruled]{algorithm2e}
\usepackage{array}
\usepackage{booktabs}
\usepackage[caption=false,font=footnotesize,labelfont=rm,textfont=rm]{subfig}
\usepackage{textcomp}
\usepackage{stfloats}
\usepackage{url}
\usepackage{verbatim}
\usepackage{graphicx}
\usepackage{cite}
\usepackage{ragged2e}
\usepackage{multirow}

\usepackage[colorlinks=true,
            linkcolor=blue,
            urlcolor=blue,
            citecolor=blue]{hyperref}

\IEEEaftertitletext{\vspace{-2\baselineskip}}

\begin{document}

\title{Prop-WRF: Proposal-Guided Wireless Radiance Field for Scalable Multi-Frequency Outdoor Channel Frequency Response Prediction
}
\author{Yulin Fu, Jiancun Fan,~\IEEEmembership{Senior Member,~IEEE}, Shiyu Zhai, Zhibo Duan, and Jie Luo
\thanks{This work was supported by the National Natural Science Foundation of China under Grants 62471381 and 62601633.

Yulin Fu, Jiancun Fan, Shiyu Zhai, Zhibo Duan, and Jie Luo are with the School of Information and Communications Engineering, Xi’an Jiaotong University, Xi’an 710049, China (Corresponding author: Jiancun Fan, e-mail: fanjc0114@gmail.com).}
}



\maketitle

\begin{abstract}
Sparse-directional wireless radiance fields (WRFs) enable complex channel frequency response (CFR) prediction from receiver locations and multipath direction priors. However, in large-scale outdoor environments, long sampling intervals make it difficult to concentrate a limited sampling budget around valid virtual anchors (VAs), degrading prediction accuracy and training stability. This paper proposes Prop-WRF, a proposal-network-guided WRF for CFR prediction across multiple subcarriers. A lightweight proposal network learns VA occupancy densities through online distillation from the main network and iteratively guides resampling toward likely VA locations. The main network then predicts complex radiance over the refined intervals and synthesizes CFRs using an analytical propagation model. By introducing frequency encoding only into the radiance branch, a single model jointly learns shared VA geometry from observations across subcarriers while representing frequency-dependent complex radiance. A stage-wise hybrid normalized loss and an interval-weight regularizer further strengthen weak-channel supervision and suppress spurious VA peaks. In addition, a multipath simultaneous localization and mapping (SLAM) front end refines receiver states and path associations to construct geometrically consistent training priors. Simulations show that Prop-WRF improves CFR prediction accuracy and training stability over conventional sparse-directional WRF baselines in larger environments.
\end{abstract}


\section{Introduction}

\IEEEPARstart{L}{ocation-aware} channel prediction is becoming an important component of environment-aware wireless communications. A channel knowledge map (CKM) uses historical or simulated location-tagged channel observations to provide channel information at unmeasured locations, thereby reducing repeated channel sounding and supporting network planning, link adaptation, and other location-dependent transmission decisions~\cite{ckm_tutorial,ckm_data}. Conventional radio maps usually describe slowly varying quantities such as received power, path loss, or interference level~\cite{radio_map_survey}. In contrast, a map of the complex channel frequency response (CFR) must preserve both magnitude and phase. Because the latter changes rapidly under coherent multipath superposition, predicting CFRs over a large outdoor region is considerably more demanding than interpolating received-power measurements.

Deterministic ray tracing (RT) resolves propagation paths from explicit geometry and material models, but incurs substantial modeling and computational costs~\cite{rt_6g}. Data-driven alternatives include RadioUNet for received-power maps and pseudo RT for outdoor millimeter-wave path loss~\cite{radiounet,pseudo_rt}. Physics-informed regional predictors also combine propagation directions and analytical transmission modules to recover channel impulse responses from sparse anchor observations~\cite{pidrtn}. These approaches generally use scene descriptions, whereas our target is continuous, phase-coherent CFR prediction from sparse wireless observations.

\begin{figure*}[t]
    \centering
    \includegraphics[width=\linewidth]{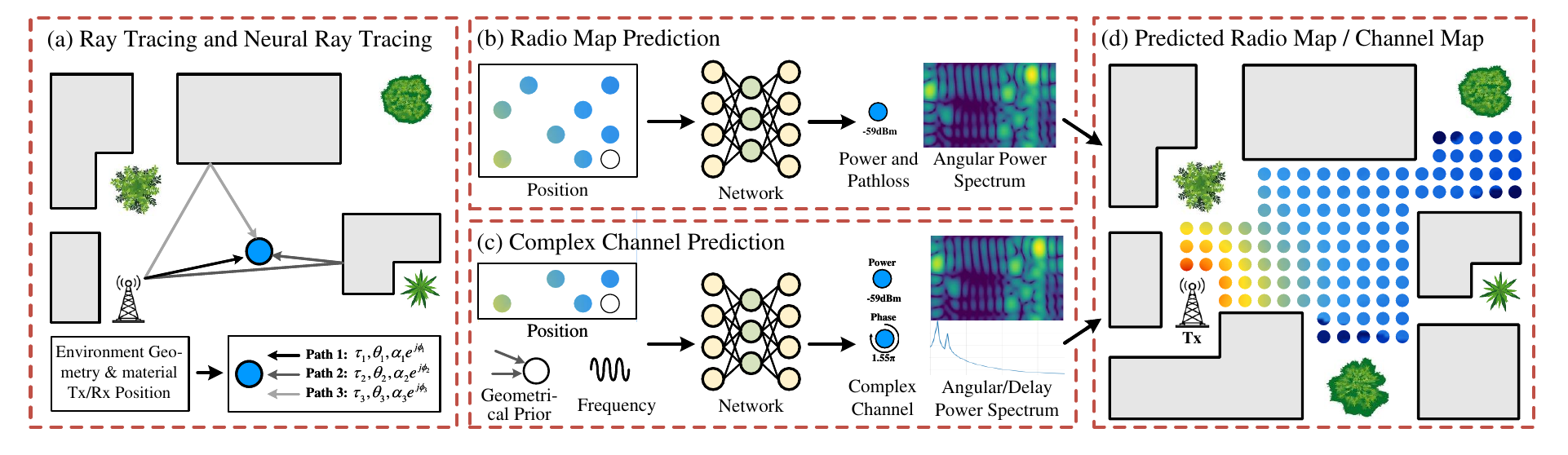}
    \caption{Conceptual comparison of location-dependent channel prediction approaches: (a) deterministic or neural ray tracing, (b) learning based radio map regression, (c) sparse-directional WRF for complex CFR prediction, and (d) spatial variation schematic diagram of the predicted power map and CFR. Prop-WRF targets efficient VA localization along long candidate rays in large outdoor regions.}
    \label{fig_intro}
\end{figure*}

Neural field representations offer another route to site-specific channel modeling. Wireless radiance fields (WRFs), inspired by the neural radiance field (NeRF) representation in computer vision~\cite{nerf}, embed the site-specific propagation field in learnable continuous functions. NeRF$^2$ established an omnidirectional WRF that synthesizes spatial radio spectra by accumulating learned attenuation and radiation along sampled rays~\cite{nerf2}. NeWRF subsequently introduced a sparse-directional formulation: given a receiver location and a small set of dominant directions of arrival (DoAs), it predicts complex directional channel components and coherently combines them into the CFR~\cite{newrf}. This formulation avoids densely sampling directions with negligible received energy and is particularly attractive when dominant multipath parameters can be estimated from channel measurements. However, such sparse methods mainly focus on spatial prediction under a fixed frequency setting~\cite{newrf,mip-newrf}. This formulation limits their direct application to multicarrier systems, where complex channel responses across multiple subcarriers are required for transmission design. The main distinctions among these channel-prediction paradigms and corresponding targets are illustrated in Fig.~\ref{fig_intro}.

Recent WRF studies have mainly improved representation or rendering efficiency. WRF-GS represents virtual anchors (VAs) and propagation using 3D Gaussian primitives and electromagnetic splatting~\cite{wrf-gs}; VoxelRF replaces repeated deep-multilayer perceptron (MLP) evaluations with an explicit voxel grid and shallow networks~\cite{voxelrf}; F$^4$-CKM introduces a spatial--frequency-aware radiator representation for fast CKM construction~\cite{f4-ckm}; and GSRF uses complex-valued Gaussians and a directional basis to retain phase-dependent RF radiance~\cite{gsrf}. These advances show that the sampling and representation cost of a WRF is as important as its network capacity. Nevertheless, less attention has been paid to the sampling failure that arises when a sparse-directional WRF is extended from a bounded indoor scene to a large outdoor region. Along a long candidate ray, a physically valid VA occupies only a narrow portion of the search interval. Uniform coarse sampling can therefore miss the useful support, whereas repeatedly querying a full CFR-prediction network merely to refine the sampling distribution incurs a high computational cost.

Higher positional encoding (PE) resolution alone does not resolve this sampling problem. Mip-NeWRF uses scale-consistent hybrid PE but still queries the channel prediction network at both coarse and fine stages~\cite{mip-newrf,mip-nerf}. In visual rendering, DONeRF concentrates samples using a depth oracle~\cite{donerf}, while mip-NeRF~360 distills a lightweight proposal MLP from the radiance network and regularizes interval distortion~\cite{mip_nerf360}. Adapting these designs to wireless channels requires care: CFR labels supervise coherent multipath sums rather than individual rays, and small distance errors can substantially rotate phase. Moreover, analytical attenuation, delay, and phase must use metric path lengths even when network coordinates are normalized. These constraints motivate precise VA localization over long physical sampling intervals.

In addition, reliable multipath priors are essential for constructing sparse-directional WRF training sets. Existing studies commonly assume that the receiver locations and associated DoA information are already available~\cite{newrf,mip-newrf}. In practice, these priors must be extracted from channel measurements collected along receiver trajectories~\cite{mmwave_slam,ref-slam-3}. Although high-resolution estimators can accurately recover individual path parameters~\cite{doa_sbl,para_esti_1,para_esti_2}, their location-wise estimates remain susceptible to noise, missed or false detections, and ambiguous path associations across time~\cite{bpslam,ref-slam-1}. Multipath SLAM exploits temporal and geometric consistency to jointly refine receiver states, VA states, and path associations~\cite{channel_slam,bpslam}. Using multipath SLAM as a front end therefore provides a natural but largely unexplored bridge from trajectory-based wireless measurements to a geometrically consistent WRF training set.

We therefore propose Prop-WRF, a proposal-network-guided WRF for outdoor CFR prediction across multiple subcarriers. Its main contributions are:
\begin{itemize}

\item We develop Prop-WRF, a proposal-guided VA sampling architecture for sparse-directional WRFs in large-scale outdoor regions. Repeated sampling passes are assigned to a lightweight proposal network, whereas complex radiance prediction and CFR synthesis are reserved for the main network after the samples have been concentrated around likely VAs. The proposal distribution is learned through online distillation from detached main-network histograms, requiring neither additional VA labels nor path-wise CFR supervision.

\item We introduce a frequency-conditioned, physics-aware formulation for phase-coherent CFR reconstruction across multiple subcarriers. Prop-WRF decouples frequency-shared VA geometry from frequency-dependent complex radiance, allowing a single model to reuse geometric support across frequencies. Analytical free-space propagation and surface-interaction effects are retained in the channel synthesis module to reduce the learning burden. Moreover, a stage-wise hybrid of globally and sample-wise normalized errors progressively emphasizes weak-channel locations, while an additional interval-weight regularizer encourages compact VA locations and suppresses spurious peaks.

\item We establish and validate a practical prior-acquisition pipeline for constructing sparse-directional WRF training datasets from trajectory-based wireless measurements. Multipath SLAM jointly refines the receiver states and associates the estimated DoA and time-of-arrival (ToA) measurements, providing spatially consistent geometric priors for WRF training. Comprehensive experiments under large-scale environments, single- and multi-frequency prediction, and imperfect-prior conditions verify the effectiveness and robustness of the proposed components, and demonstrate that Prop-WRF achieves higher CFR-prediction accuracy than representative WRF baselines.
\end{itemize}

The remainder of this paper is organized as follows. Section~II introduces the received-signal and multipath-SLAM models. Section~III presents the proposal and main networks, the channel synthesis procedure, and the training objectives. Section~IV describes the datasets and evaluates Prop-WRF. Finally, Section~V concludes the paper.

\section{System Model}
\subsection{Received Signal Model}
In a typical channel prediction task consisting of a single antenna receiving link, assume the spatial location of the receiver at discrete time $t$ is $\mathbf{x}_t$. The channel frequency response can then be represented as:
\begin{equation}\label{eq2-1}
    H_t(f)=\sum_{l=1}^{L_t} {\alpha}_{t,l}(f) e^{-j2\pi f {\tau}_{t,l}},
\end{equation}
where $L_t$ is the path number, $\mathcal{P}\!_t\!=\! \{{\alpha}_{t,l}(f), {\tau}_{t,l}, {\theta}_{t,l}, {\varphi}_{t,l} \}_{l=1}^{L_t}$ is the associated information denoting path gain, delay, zenith and azimuth angle of each path, and $f$ is carrier frequency. Note that for multi-antenna receiver Eq.~\eqref{eq2-1} also holds by considering only the first antenna. We consider a set of $K$ subcarrier frequencies $\{f_k\}_{k=1}^{K}$. Since the considered bandwidth is small, the propagation geometry is assumed to be shared across these frequencies. In contrast, the complex radiation coefficient and the propagation phase remain frequency dependent.

In general propagation environments, the dominant channel components can often be characterized by the LoS path and several dominant specular reflection and transmission paths \cite{model1}. Accordingly, scattering and diffraction effects are omitted, affects only the channel synthesis stage, while they may still be learned implicitly by the neural network from training observations. Under this assumption, the path gain can be decomposed as ${\alpha}_{t,l}(f_k)\!=\!a_{t,l} \zeta_{t,l} \varsigma_{t,l} $. The free space propagation loss $a_{t,l}$ is calculated by $a_{t,l}\!=\!(4\pi {\tau}_{t,l} f)^{-1}$. The surface interaction loss is obtained by the modulus product of all reflection and transmission losses on the $l$-th path, i.e. $\zeta_{l,t}\!=\!\prod _j |\zeta_{l,t,j}|$, where the $j$-th interaction coefficient is calculated through single-layer slab model \cite{ITU_P2040}:
\begin{equation}\label{eq2-2}
\begin{aligned}
    \zeta_{l,t,j}^{(\mathrm{ref})} &=\frac{r_{\perp}(1-e^{-j2q})}{1-r_{\perp}^2e^{-j2q}}, \quad\text{for reflection},
    \\
    \zeta_{l,t,j}^{(\mathrm{tra})} &=\frac{(1-r_{\perp}^2)e^{-jq}}{1-r_{\perp}^2e^{-j2q}}, \quad \text{for transmission},
\end{aligned}
\end{equation}
where $r\!_{\perp}$ is transverse electric (TE) polarization Fresnel reflection coefficient \cite{ITU_P2040} (as approximation of the actual value providing as an attenuation reference for network):
\begin{equation}\label{eq2-3}
\begin{aligned}
    r_\perp&=\frac{\cos(\theta_{inc})\!-\!\sqrt{\eta\!-\sin^2(\theta_{inc})}}{\cos(\theta_{inc})\!+\!\sqrt{\eta-\sin^2(\theta_{inc})}},
    \\
    t_\perp&=\frac{2\cos(\theta_{inc})}{\cos(\theta_{inc})+\sqrt{\eta-\sin^2(\theta_{inc})}},
\end{aligned}
\end{equation}
where $\theta_{inc}$ is the incident angle, $\eta$ is the complex material-dependent relative permittivity, assumed to be known a priori by existing estimation methods \cite{material1,material2,material3}. The electrical phase thickness satisfies $q\!=\!2\pi d \sqrt{\eta\!-\!\sin^2\theta_{inc}} /\lambda$, in which $d$ is slab thickness (at the reflection point) and $\lambda\!=\!c/f$ is wavelength. The residual attenuation factor $ \varsigma_{t,l} $ accounts for the interaction coefficient approximation error and other unmodeled propagation impairments such as diffuse scattering and diffraction.

\subsection{Multipath SLAM Model}
Conventional WRF datasets assume available receiver locations and noisy path parameters~\cite{newrf,mip-newrf}. For trajectory-based acquisition, we instead refine estimated DoAs and ToAs~\cite{para_esti_1,para_esti_2} using a multipath SLAM front end. Temporal and geometric consistency support joint receiver localization and path association, yielding consistent training priors~\cite{bpslam,ref-slam-1,ref-slam-2,ref-slam-3}.

The transmitter is modeled as a physical anchor (PA), and each specular reflection path is represented in terms of virtual anchors (VAs, mirror images of PA). The PA and all corresponding VAs are called potential features (PF) with state $\mathbf{m}_{k,t}\!=\![\mathbf{p}^{\mathrm{T}}_{k,t} , r_{k,t}] ^{\mathrm{T}}$, where $k$ is the PF index, $\mathbf{p}_{k,t}$ is PF's position and $r_{k,t}$ is an auxiliary variable indicating the PF's presence/absence at time $t$. The receiver state is defined as $\mathbf{s}_t\!=\![\mathbf{x}_t^\mathrm{T},\mathbf{v}_t^\mathrm{T}]^\mathrm{T}$, where $\mathbf{v}_t$ denotes the receiver velocity. For each path $l$, parameter estimation provides a measurement vector $\mathbf{z}_{t,l}\!=\![\hat{\tau}_{t,l}, \hat{\theta}_{t,l}, \hat{\varphi}_{t,l}]^\mathrm T$, consisting of the estimated ToA and DoA. This measurement is related to the corresponding PF state $\mathbf{m}_{k,t}$ and receiver state $\mathbf{s}_t$, i.e.,
\begin{equation}\label{eq2-4}
    \mathbf{z}_{t,l}=h(\mathbf{m}_{k,t},\mathbf s_t)+\mathbf n_{t,l},
\end{equation}
where the $k$-th PF is assumed to correspond to the $l$-th path, $h(\cdot)$ denotes the geometric measurement model and $\mathbf n_{t,l}$ represents estimation noise.

Using all measurements $\mathbf{z}_{t,l}$, the posterior distribution is approximated to give a joint estimation results of the receiver trajectory, PF map, and path associations, i.e., $p(\mathbf{m}_{k,t},\mathbf{s}_{t},\mathbf c_{k,t,l}|\mathbf{z}_{t,l})$ for all $t\!=\!1,\ldots,T$ and $L\!=\!1,\ldots,L_t$, in which zero-one variable $c_{k,t,l}$ means whether the $k$-th PF corresponds to the $l$-th path or not at time $t$. Following \cite{bpslam}, belief propagation on a factor graph is adopted to obtain the MMSE estimates of the receiver trajectory $\hat{\mathbf{x}}_t$ and associated path parameters $\hat{\mathcal P}_t$, which is subsequently used for WRF dataset construction and network training. 


\section{Prop-WRF Framework}
\subsection{Overview}
To balance computational efficiency and physical consistency, Prop-WRF follows the sparse directional WRF framework that performs conditional modeling only along the DoA dimension for each VA. Under the VA representation, each non-line-of-sight (nLoS) component from the transmitter to the receiver can be viewed as a direct path from the corresponding VA to the receiver. Accordingly, channel synthesis is carried out in three steps: estimating the VA probability density distribution over DoA, predicting the equivalent transmitted signal at discretely sampled DoA points, and finally aggregating all samples through a physically consistent weighted combination. The first two steps are implemented by the MLP, while its output experiences an analytical synthesis module to realize the third step. 

In large-scale outdoor scenarios with sampling ranges about 100m, the conventional two stage coarse-to-fine strategy fails to reliably localize the correct VA peak. In our experiments, it consistently suffers from collapse and zero output. Although increasing the number of passes to four can alleviate this issue, the resulting performance is still mediocre, while the computational cost is substantially increased. This suggests that, under long sampling ranges, the early sampling stages are mainly used to provide a rough prior over the VA distribution, whereas the channel synthesized from such coarse samples contributes little to network training.

\begin{figure*}[t]
    \centering
    \includegraphics[width=\textwidth]{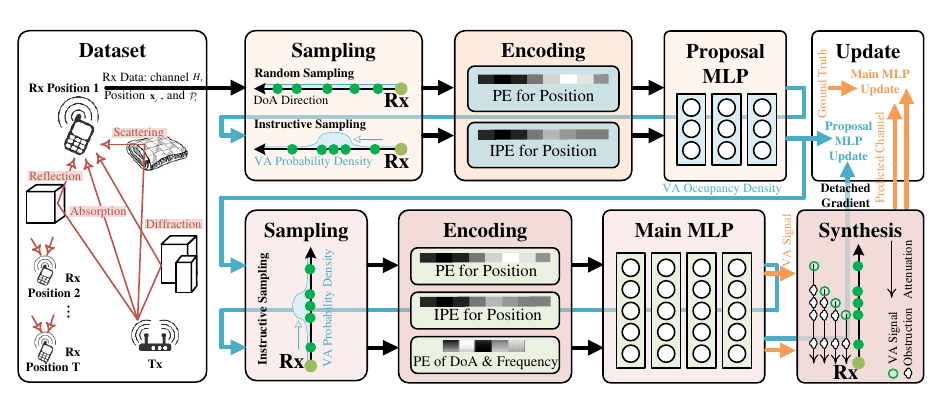}
    \caption{Overview of prop-WRF framework. The blue and orange arrows indicate the flow of VA occupancy density and equivalent VA signal, respectively. Sampling and encoding are adopted firstly to extract the effective information at the receiver, and the CFR is synthesized using the MLP network predictions. In large scale environment, prop-WRF achieves robust and efficient prediction by proposal MLP to quickly converge VA probability density in DoA.}
    \label{fig_a}
\end{figure*}

As illustrated in Fig.~\ref{fig_a}, we therefore introduce a lightweight proposal network to learn a coarse VA distribution prior along a DoA-conditioned ray by distilling it from the main network. Based on this prior, the main MLP performs importance sampling to obtain a more accurate VA representation in an efficient manner. In our implementation, only one proposal MLP and one main MLP are used, with setting $T_p\!=\!2$ and $T_m\!=\!2$ passes, respectively, to balance cost between sampling network prediction rounds.

Following~\cite{mip-newrf}, we encode conical frustums along each candidate DoA using hybrid PE and integrated PE (IPE). The main network predicts interval weights and complex radiance for analytical CFR synthesis.

\subsection{Proposal Network Design}

\subsubsection{VA Occupancy Density Prediction}
Consider an arbitrary receiver location $\mathbf{x}_t$, where the associated multipath set $\mathcal{P}_t$ is assumed to be already estimated. The proposal MLP is employed to predict the existence probability distribution of VAs over the DoA space.

For notational simplicity, denote the receiver position by $\mathbf{o}$ and a candidate unit DoA direction by $\hat{\mathbf{d}}$. To sample $m$ candidate frustum intervals that may contain VAs along direction $\hat{\mathbf{d}}$, we first construct a cone originating from $\mathbf{o}$ with height $d_f$ and base radius $d_f\dot r$, where $\dot r$ is a configurable scaling parameter controlling the cone aperture. Next, $m\!+\!1$ stratified samples are generated within the range $[0,d_f]$, where the $i_s$-th sample distance is drawn from $d_{i_s} \!\in\! [{({i_s}\!\!-\!\!1\!)d\!_f}/(\!m\!+\!1\!),{{i_s}d\!_f}/(\!m\!+\!1\!)]$ with ${i_s}=1,\ldots,m\!+\!1$, while the first two sample locations are fixed as $d_1\!\!=\!\!10^{-3}$ and $d_2\!\!=\!\!2\!\!\times\!\!10^{-3}$ to avoid numerical singularities. This design also fixes the location of the first sampling interval, which serves as a NeRF-style reference origin that aggregates the signals propagated from VAs associated with all sampled intervals. The $i$-th frustum interval is then defined as:
\begin{equation}\label{eq3-0}
\Gamma\!_i(\mathbf{x})\! =\!\! \big(d_{i} \!< \! \hat{\mathbf{d}}^{\mathrm{T}}\!(\mathbf{x}\!-\!\mathbf{o}) \!<\! d_{i+1}\big) 
 \!\wedge\! \bigg(\!\frac{\hat{\mathbf{d}}^{\mathrm{T}}\!(\mathbf{x}\!-\!\mathbf{o})}{\left\|\mathbf{x}\!-\!\mathbf{o}\right\|_{2}}\!\!>\!\!\frac{1}{\sqrt{1\!+\!\dot{r}^{2}}}\!\bigg),
\end{equation}
where $i\!=\!1,\!\ldots\!,\!m$.

After sampling, each $i$-th frustum interval is encoded using a hybrid scheme that combines PE and IPE. This design preserves estimation accuracy while improving robustness to sampling uncertainty \cite{mip-newrf}. For the PE branch, we apply standard PE to the first-order expectation of the position within $\Gamma_i(\mathbf{x})$, i.e.,
\begin{equation}\label{eq3-1}
    \mathbf{x}=\mathbf{o}+\frac{3\left(d_{i+1}^4-d_i^4\right)}{4\left(d_{i+1}^3-d_i^3\right)}\hat{\mathbf{d}}.
\end{equation}

The PE map is defined as:
\begin{equation}\label{eq3-2}
    \gamma_{\mathrm{pe}}(\Gamma_i,\mathbf{L})=
    \mathbf{x}'
    \cup
    \left\{
    \begin{bmatrix}
    \sin\!\left(2^{\boldsymbol\ell}\pi \mathbf{x}'\right) \\
    \cos\!\left(2^{\boldsymbol\ell}\pi \mathbf{x}'\right)
    \end{bmatrix}
    \right\}_{\boldsymbol{\ell}=\mathbf{0}}^{\mathbf{L} \!- \!\mathbf{1}},
\end{equation}
where $\mathbf{x}'\!\!=\!(\mathbf{x}\! - \!\mathbf{x}_{\min})/(\mathbf{x}_{\max} \!-\! \mathbf{x}_{\min})$ denotes the normalized coordinate, $\boldsymbol\ell \!=\![l_x,l_y,l_z]$ is the dimension variable, and $\mathbf{L}\!=\![L_x,L_y,L_z]$ is the corresponding maximum encoding order.

For the IPE branch, we integrate the PE features over the distribution induced by $\Gamma_i(\mathbf{x})$, yielding:
\begin{equation}\label{eq3-3}
\gamma_{\mathrm{ipe}}(\Gamma_i,\!\mathbf{L}) \!\!=\!\!
\left\{\!\!
\begin{bmatrix}
\sin\!\left(2^{\boldsymbol\ell}\pi \boldsymbol{\mu}'\right)\!
\exp\!\left(\!-\pi^2 2^{2\boldsymbol\ell \!-\! 1} \! \operatorname{diag}(\boldsymbol{\Sigma}\!')\!\right)\!\! \\
\cos\!\left(2^{\boldsymbol\ell}\pi \boldsymbol{\mu}'\right)\!
\exp\!\left(\!-\pi^2 2^{2\boldsymbol\ell \!-\! 1} \! \operatorname{diag}(\boldsymbol{\Sigma}\!')\!\right)\!\!
\end{bmatrix}\!\!
\right\}_{\boldsymbol\ell=0}^{\mathbf{L} \!-\! \mathbf{1}},
\end{equation}
where normalized coordinate $\boldsymbol{\mu}'\!\!=\!({\boldsymbol{\mu}\!-\!\mathbf{x}_{\min}}) / ({\mathbf{x}_{\max}\!-\!\mathbf{x}_{\min}})$, and $
\operatorname{diag} (\boldsymbol{\Sigma}')\! \!=\! {[\Sigma_{xx},\Sigma_{yy},\Sigma_{zz}]^\top}/{(\mathbf{x}_{\max}\!-\!\mathbf{x}_{\min})^2}$. Here, $\boldsymbol{\mu}$ and $\boldsymbol{\Sigma}$ denote the mean and covariance of the position distribution within the target sampling interval $\Gamma_i(\mathbf{x})$, respectively. The mean satisfies Eq.~\eqref{eq3-1} and the covariance is given by:
\begin{equation}\label{eq3-4}
\begin{aligned}
\boldsymbol{\Sigma}
&=
\bigg[
\frac{3\left(d_{i+1}^5-d_i^5\right)}{5\left(d_{i+1}^3-d_i^3\right)}
-\frac{9\left(d_{i+1}^4-d_i^4\right)^2}{16\left(d_{i+1}^3-d_i^3\right)^2}
\bigg]\hat{\mathbf{d}}\hat{\mathbf{d}}^\top \\
&\quad+
\frac{3\dot{r}^{\,2}\left(d_{i+1}^5-d_i^5\right)}{20\left(d_{i+1}^3-d_i^3\right)}
\big(\mathbf{I}-\hat{\mathbf{d}}\hat{\mathbf{d}}^\top\big).
\end{aligned}
\end{equation}

The proposal network aims to predict an occupancy density $\sigma_i^{\mathrm{(p)}}$ that indicates the VA existence within each sampled interval, i.e.,
\begin{equation}\label{eq3-5}
    f_{\text{prop}}\big(\gamma_{\mathrm{pe}}(\Gamma_i,\mathbf{L}),\gamma_{\mathrm{ipe}}(\Gamma_i,\mathbf{L})\big)\rightarrow \sigma_i^{\mathrm{(p)}}.
\end{equation}
Unlike \cite{mip_nerf360}, we directly use the predicted occupancy density rather than the compositing weights (further synthesized by Eq.~\eqref{eq3-13}). Over a long sampling range of approximately 100 m, even a small occupancy density at an early sample can substantially attenuate subsequent samples through accumulated transmittance. During proposal-network training, this effect may create a strong positive-feedback loop that concentrates the predicted VA distribution near the first few samples and impedes optimization. The occupancy density is therefore a more suitable equivalent sampling-weight density representation.

In practice, the proposal network, which is implemented as a lightweight MLP with 4 layers, 64 hidden units per layer, and ReLU activations, yields low inference overhead.

\subsubsection{Importance Resampling} \label{resampling}

During the resampling stage, importance sampling is performed according to the prior VA existence equivalent weight density. After obtaining the interval-wise equivalent weight $\{\sigma_i^{\mathrm{(p)}}\}_{i=1}^{m}$ along the target DoA $\hat{\mathbf{d}}$, we apply a local smoothing operation to alleviate sensitivity to overly narrow high-confidence intervals and to preserve a broader support for resampling:
\begin{equation}\label{eq3-6}
    \sigma_i'  =\frac{1}{2}\Big(\max(\sigma^{\mathrm{(p)}}_{i-1},\sigma^{\mathrm{(p)}}_i)+\max(\sigma^{\mathrm{(p)}}_i,\sigma^{\mathrm{(p)}}_{i+1})\Big).
\end{equation}
Assuming a uniform axial density distribution within each sampling interval, the probability density along the sampling ray is defined as:
\begin{equation}\begin{aligned}\label{eq3-7}
    &f(d_j)={\sigma_i'}+\epsilon, & &d_j\!\in\![d_i,d_{i+1})\\
    &f(d_j)=\epsilon, & &d_j\!\in\!(0,d_1)\cup[d_{m+1},f_f)
\end{aligned}\end{equation}
where $\epsilon$ is a predefined density compensation term that ensures a non-zero sampling probability everywhere along the ray and provides recovery capability when the resampling focus is misplaced. Accordingly, the cumulative distribution function (CDF) of occupancy density can be written as:
\begin{equation}\label{eq3-8}
    F(d_j)=\frac{1}{S}\left[\sum_{k=1}^{i-1} {(d_{k+1}\!-\!d_k)}\sigma_k'+ ({d_j \!-\! d_i})\sigma_i'  +\epsilon d_j \right],
\end{equation}
where $d_j\!\in\![d_i,d_{i+1})$ and $S \!=\! [\sum_{j=1}^{m}{(d_{j+1} \!-\! d_j)}{\sigma_j'}] \!+\! d_f \epsilon$ is the normalization factor. And specifically, $F(d_j)\!=\!\epsilon d_j /S$ when $d_j\!\in\!(0,d_1)$ and $F(d_j)\!=\!1\!-\! \epsilon (d_f\!-\!d_j)/S$ when $d_j\!\in\![d_{m+1},f_f)$.

We then draw $m\!+\!1$ i.i.d. samples $u_i \!\sim\! \mathcal{U}(0,1)$ and obtain the resampled distances by inverse transform sampling, i.e., $d_i^{(2)'} \!\!=\! F^{-1}(u_i)$. After sorting $\{d_i^{(2)'}\}$ in ascending order, we again set $d_1^{(2)} \! =\! 10^{-3}$ and $d_2^{(2)}\! =\! 2\! \times\!  10^{-3}$ to avoid singularities, resulting in the final importance-resampled set $\{d_i^{(2)}\}_{i=1}^{m+1}$. The corresponding sampling intervals and VA existence are obtained in the same manner as in Eq.~\eqref{eq3-0} and \eqref{eq3-5}. The final output of the proposal MLP to the main MLP in our architecture is the frustum $\boldsymbol{\Gamma}^{(2)} \!=\!  \{\Gamma_i^{(2)}\}_{i=1}^{m}$ and the equivalent weight density $\boldsymbol{\sigma}^{(2)} \!=\! \{\sigma_i^{(2)}\}_{i=1}^{m}$ associated with the DoA.

\subsubsection{Proposal Network Training}
\begin{figure}[t]
    \centering
    \includegraphics[width=\linewidth]{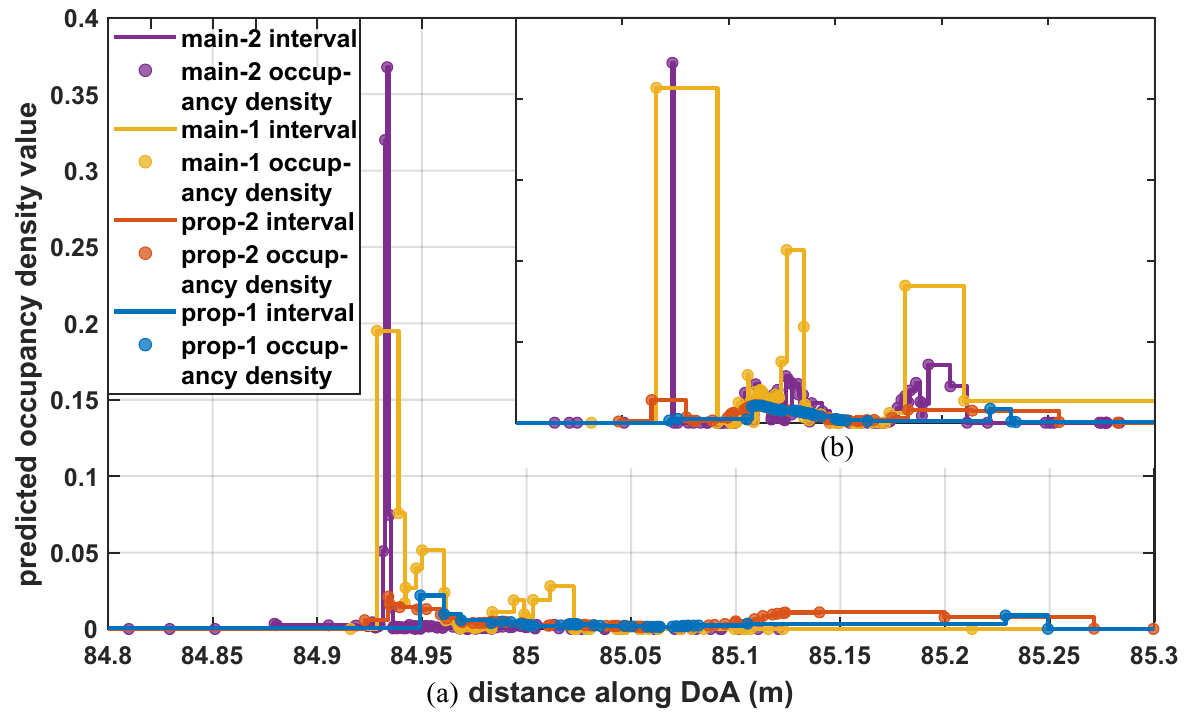}
    \caption{Schematic of the sampling interval occupancy densities prediction results by proposal network and main network, (a) the results at 10k training iterations along a certain ray, (b) the results at 1k iterations along the same ray. Under long distance sampling, Prop-WRF can gradually focus the sampling to the vicinity of the target VA distance and suppress potential multiple erroneous VAs originally predicted by the network during training.}
    \label{fig_b}
\end{figure}

The objective of the proposal MLP is to provide a computationally efficient prior over the spatial distribution of VAs, as illustrated in Fig.~\ref{fig_b}. Consequently, it does not participate in the channel prediction process and cannot be directly supervised using the CFR labels available in the dataset.

The final main-network pass (Section~\ref{main-net}) produces intervals and occupancy densities $(\boldsymbol{\Gamma}^{(4)},\boldsymbol{\sigma}^{(4)})$ that supervise the proposal histogram $(\boldsymbol{\Gamma}^{(2)},\boldsymbol{\sigma}^{(2)})$ through online distillation. Since the interval partitions differ, we compare their integrated masses using an overlap-based bound inspired by~\cite{mip_nerf360}.

Although resampling assumes uniform density within each interval, the true within-interval distribution is unknown. We therefore require the proposal mass over intervals overlapping a main-network interval to cover that interval's target mass:
\begin{equation}\begin{aligned}
    &\mathrm{Sup} \big(\boldsymbol{\Gamma}^{(2)}, \boldsymbol{\sigma}^{(2)},{\Gamma}^{(4)}_i\big)
    \\&= \sum_{k: {\Gamma}^{(2)}_k\cap{\Gamma}^{(4)}_i \neq \varnothing}  {\sigma}^{(2)}_k (d_{k+1} \!-\! d_k) \geq {\sigma}^{(4)}_i (d_{i+1} \!-\! d_i).
\end{aligned}\label{eq3-9}
\end{equation}

In general, owing to the lack of prior knowledge, the proposal MLP tends to produce relatively uniform sampling intervals and overestimates their equivalent weights. In contrast, intervals in vicinity of the target VAs are progressively refined by the main MLP, resulting in denser sampling and higher equivalent weights (see Fig.~\ref{fig_b}). Therefore, we penalize only the inconsistency caused by underestimating the probability mass of the proposal histogram:
\begin{equation}\begin{aligned}\label{eq3-10}
    &\mathcal{L}_{\mathrm{prop}} \big(\boldsymbol{\Gamma}^{(2)}\!,  \boldsymbol{\sigma}^{(2)} \!, \boldsymbol{\Gamma}^{(4)}\!, \boldsymbol{\sigma} ^{(4)} \big) \!=\! \sum_i \!\frac{1}{(d_{i+1} \!-\! d_i)\sigma_i^{(4)} \!+\!\epsilon_{\mathrm{loss}}}\\  &    \times\! \Big[\max\big(0, \sigma_i^{(4)} - \mathrm{Sup} (\boldsymbol{\Gamma}^{(2)}, \boldsymbol{\sigma}^{(2)},{\Gamma}^{(4)}_i) \big)\Big]^2,
\end{aligned}\end{equation}
in which the normalization factor $1/{\sigma_i^{(4)} \!+\! \epsilon_{\mathrm{loss}}}$ prevents the loss from vanishing to zero when the proposal probability mass $\mathrm{Sup}(\boldsymbol{\Gamma}^{(2)},  \boldsymbol{\sigma}^{(2)},\Gamma_i^{(4)})$ approaches zero, thereby improving optimization stability. During training, the equivalent weights $ \boldsymbol{\sigma}^{(4)}$ are detached from the computational graph before evaluating \eqref{eq3-10}. Otherwise, the coupling between the proposal and main networks may drive the optimization toward undesirable local minima associated with degraded channel prediction performance. Since the first- and second-stage proposal networks produce different interval partitions and equivalent weights $(\boldsymbol{\Gamma}^{(1)}, \boldsymbol{\sigma}^{(1)})$ and $(\boldsymbol{\Gamma}^{(2)}, \boldsymbol{\sigma}^{(2)})$, the proposal loss in \eqref{eq3-10} is applied to both stages independently.

\subsection{Main Network Design}

\begin{figure*}[t]
    \centering
    \includegraphics[width=\linewidth]{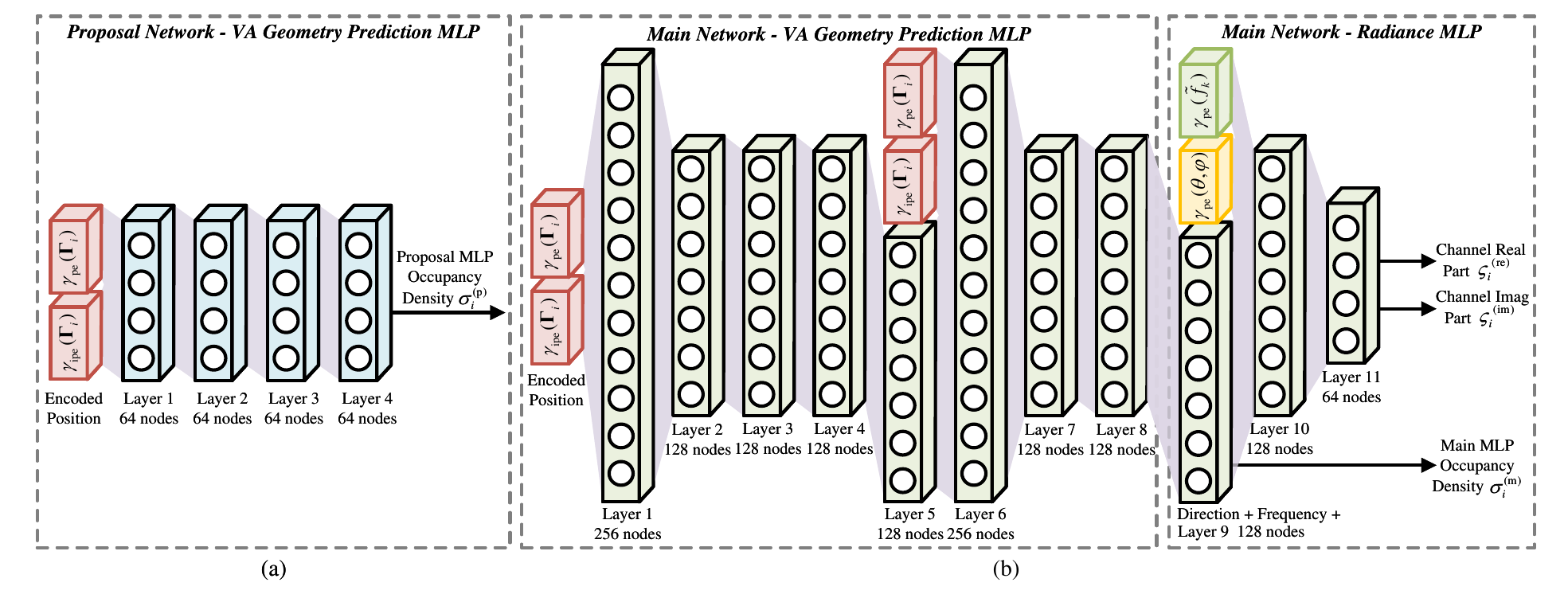}
    \caption{Network architecture of (a) proposal MLP, and (b) main MLP.}
    \label{fig_net_arch}
\end{figure*}

\subsubsection{CFR Prediction}\label{main-net}
The main network follows four sequential steps, namely resampling, feature encoding, network prediction, and channel synthesis. The resampling procedure is identical to that described in subsection~\ref{resampling}, where the sampling intervals are updated according to the occupancy density distribution predicted in the previous stage. For example, when entering the first main-network stage, importance resampling is performed based on $(\boldsymbol{\Gamma}^{(2)}, \boldsymbol{\sigma}^{(2)})$.

Unlike the proposal network, the main network predicts not only the VA occupancy density within each sampling interval using the VA geometry prediction MLP, but also the corresponding equivalent complex signal coefficient emitted toward the receiver using the radiance MLP, as shown in Fig.~\ref{fig_net_arch}. Two additional inputs, namely the direction and frequency encodings, are fed only into the radiance MLP. This design is motivated by the fact that VA geometry is an environment-dependent property shared across propagation directions and subcarrier frequencies, whereas the equivalent radiance varies with both direction and frequency.

The incident direction is encoded using the zenith and azimuth angles, denoted by $\theta$ and $\varphi$, respectively, to distinguish rays passing through identical spatial locations but having different propagation directions. Since the directional dependence varies smoothly, low-dimensional PE is adopted, i.e., $\gamma_{\mathrm{pe}}(\theta,L_\theta)$ and $\gamma_{\mathrm{pe}}(\varphi,L_\varphi)$ with small $L_\theta$ and $L_\varphi$. Similarly, the encoded subcarrier frequency $\gamma_{\mathrm{pe}}(\tilde f_k,L_f)$ is introduced with $L_f\!=\!4$, where the normalized frequency is calculated as:
\begin{equation}\begin{aligned}
    \tilde{f}_k=2\frac{f_k-f_{\mathrm{min}}}{f_{\mathrm{max}}-f_{\mathrm{min}}}-1,
\end{aligned}\label{eq3-fre}
\end{equation}
and $f_k$ denotes the actual frequency of the $k$-th subcarrier. The rapid phase rotation caused by propagation is explicitly retained in the analytical channel synthesis model. Consequently, the radiance MLP only needs to learn the residual frequency-dependent variations, such as those arising from interaction-coefficient variations not fully captured by the analytical model.

The main network is formulated as:
\begin{equation}\begin{aligned}
    f_{\mathrm{main}}
    \Big(&\gamma_{\mathrm{pe}}(\Gamma_i,\mathbf{L}),
    \gamma_{\mathrm{ipe}}(\Gamma_i,\mathbf{L}),
    \gamma_{\mathrm{pe}}(\theta,L_\theta),    
    \gamma_{\mathrm{pe}}(\varphi,L_\varphi),
    \\&
    \gamma_{\mathrm{pe}}(\tilde f_k,L_f),
    \Big)    
    \rightarrow \left(\sigma^{\mathrm{(m)}}_i,\varsigma_i^{(\mathrm{re})} (f_k), \varsigma_i^{(\mathrm{im})} (f_k) \right),
\end{aligned}\label{eq3-11}
\end{equation}
where $\sigma_i^{\mathrm{(m)}}$ denotes the occupancy density of the $i$-th sampling interval, and $\varsigma_i \!=\! \varsigma_i^{(\mathrm{re})} (f_k)\!+\! j\varsigma_i^{(\mathrm{im})}(f_k)$ is the equivalent complex signal coefficient predicted separately through its real and imaginary components. The calculation of NeRF-style volumetric rendering compositing weight is defined as $w_i^{\mathrm{(\cdot)}}$, following:
\begin{equation}\begin{aligned}\label{eq3-13}
    w_i^{\mathrm{(\cdot)}} = \exp \Big(-\sum _{j=1}^{i-1} \sigma_j^{\mathrm{(\cdot)}}\delta_j \Big) \big(1-\exp(-\sigma_i^{\mathrm{(\cdot)}}\delta_i) \big),
\end{aligned}\end{equation}
where $\delta_i=d_{i+1}-d_i$. The opacity term favors occupied intervals, while accumulated transmittance attenuates intervals behind them. Multiple peaks may nevertheless persist; the regularizer introduced below encourages compact VA support.

Without loss of generality, let the number of main MLP stages be $T_m\! =\!2$. After the second stage, the predicted intervals and equivalent weight densities $(\boldsymbol{\Gamma}^{(4)}, \boldsymbol{\sigma}^{(4)})$ are used both to supervise the proposal network through the proposal loss and, together with the corresponding frequency-dependent complex coefficients $\boldsymbol{\varsigma}^{(4)} \!=\! \{\varsigma_i^{(4)}\}_{i=1}^{m}$, to synthesize the wireless channel.

The main MLP has 11 layers, 64--256 hidden units per layer, ReLU activations, and skip connections (Fig.~\ref{fig_net_arch}). Its geometry branch predicts occupancy density, while its radiance branch predicts the complex signal coefficient.

Within the proposed WRF, the CFR contribution of the $l$-th propagation path is obtained by coherently superposing the equivalent signals emitted from all sampled intervals along the corresponding direction:
\begin{equation}\begin{aligned}\label{eq3-12}
    \hat H_l=\sum_{i=1}^m \frac{c}{4\pi f d_{\mathrm{eff},i}} \zeta_i(f_k) \varsigma_i \exp(-j\frac{2\pi f d_{\mathrm{eff},i}}c) w_i,
\end{aligned}\end{equation}
where the effective sampling distance $d_{\mathrm{eff},i}$ is identical to the expectation used in Eq.~\eqref{eq3-1}, i.e., $d_{\mathrm{eff},i} = {3\left(d_{i+1}^4 \!-\! d_i^4\right)}/{4\left(d_{i+1}^3 \!-\! d_i^3\right)}$; $\zeta_i(f_k)$ denotes the surface interaction coefficient computed from the incident angle according to Eq.~\eqref{eq2-2}; $c$ is the speed of light; and $w_i$ is the NeRF-style compositing weight.

Equation~\eqref{eq3-12} explicitly compensates for both the free-space propagation loss and the surface interaction loss, thereby reducing the learning burden of the neural network. The final predicted CFR is obtained by coherently summing the contributions from all estimated DoAs:
\begin{equation}\begin{aligned}\label{eq3-14}
    \hat H=\sum_{l=1}^{L_d} \mathbf{w}^{*\top}_{\mathrm{ant}} \mathbf{c}_{\mathrm{ant}} \hat H_l,
\end{aligned}\end{equation}
where $L_d$ denotes the number of estimated DoAs, which is not necessarily equal to the true number of propagation paths $L_t$, $\mathbf{w}^{*}_{\mathrm{ant}}$ and $\mathbf{c}_{\mathrm{ant}}$ are the beamforming weight vector and antenna phase coefficient, respectively, determined by different antenna configurations. Since this paper considers only a single antenna receiver link, we set $\mathbf{w}_{\mathrm{ant}}\!=\!\mathbf{c}_{\mathrm{ant}} \!=\!1$.

\subsubsection{Main Network Training}
Each sample receiver is indexed by a location--frequency pair $(\mathbf{x}_t,f_k)$, then the main network synthesizes the complex value channel $H_l$ along the target propagation direction. However, individual location signals are indistinguishable in the training dataset labels. Therefore, the channels of all paths arriving at each receiver are predicted respectively, then summed and supervised by the corresponding ground truth channel. During multi-frequency training, the same spatial split is used at all frequencies to prevent location leakage. In each training iteration, a fixed number of receiver locations, denoted by $N_{\mathrm{batch}}$, are randomly selected for training, whereas the actual number of training samples depends on the number of paths, characterized by their DoAs, at each location. In addition to the ground truth paths, we introduce DoAs corresponding to nonexistent paths as negative samples to balance the training process.

Since the signal power decays quadratically with propagation distance, the losses associated with weak signal locations can be overwhelmed by those associated with strong signals, resulting in inaccurate predictions for weak channels. To mitigate this issue, the normalized mean squared error (NMSE) is commonly adopted in WRF~\cite{newrf, mip-newrf}:
\begin{equation}\begin{aligned}\label{eq3-15}
    \mathcal{L}_\mathrm{m1}({H},\hat {H}) =
    \frac{ \sum_{\mathrm{Rx\ batch}} \big\| H - \hat H \big\|^2}{ \sum_{\mathrm{Rx\ batch}} \big\| H \big\|^2}.
\end{aligned}\end{equation}
Compared with the conventional MSE, this formulation is more robust to variations in channel magnitude. Nevertheless, the losses of weak channel samples within the same batch may still contribute insufficiently to the supervision.

To address this limitation, we propose two normalized weighted MSE formulations. The first is the magnitude-weighted normalized MSE:
\begin{equation}\begin{aligned}\label{eq3-16}
    \mathcal{L}_\mathrm{m2}({H},\hat {H}) =
    \frac{  \sum_{\mathrm{Rx\ batch}} \frac{\| H - \hat H \|^2}{\| H \|}}{ \sum_{\mathrm{Rx\ batch}} \big\| H \big\|}.
\end{aligned}\end{equation}
In addition to global normalization, this loss assigns each sample a weight of $1/\left\|H\right\|$, thereby placing greater emphasis on weak channel samples.

The second formulation is the mean sample-wise normalized MSE (SNMSE):
\begin{equation}\begin{aligned}\label{eq3-17}
    \mathcal{L}_\mathrm{m3}({H},\hat {H}) =\sum_{\mathrm{Rx\ batch}} \frac{1}{N_{\mathrm{batch}}}\frac{\big\| H - \hat H \big\|^2}{\big\| H \big\|^2}
\end{aligned}\end{equation}
This formulation independently normalizes the channel prediction error at each receiver location, further reducing the model's bias toward strong channel samples. However, it may amplify noise and make the training process overly sensitive to samples with low signal-to-noise (SNR) ratios. 

We therefore adopt a stage-wise hybrid weighted MSE loss that combines the three loss terms:
\begin{equation}\begin{aligned}
\label{eq3-18}
    \mathcal{L}_{\mathrm{m}}({H}\!,\hat {H}\!,t) \!=\! c_{\mathrm{m1}}(t)\mathcal{L}_{\mathrm{m1}}    \!+\! c_{\mathrm{m2}}(t)\mathcal{L}_{\mathrm{m2}} \!+\! c_{\mathrm{m3}} (t)\mathcal{L}_{\mathrm{m3}},
\end{aligned}\end{equation}
where $t$ denotes the training iteration, $c_{\mathrm{m1}}(t)$, $c_{\mathrm{m2}}(t)$, and $c_{\mathrm{m3}}(t)$ are stage-dependent weighting coefficients. A predefined iteration threshold $T_{\mathrm{w}}$ is used to divide the training process into a warm-up stage and a main optimization stage. The loss coefficient schedule is defined as
\begin{equation}\begin{aligned}
\label{eq3-19}
\bigl(c_{\mathrm{m1}}(t), c_{\mathrm{m2}}(t), c_{\mathrm{m3}}(t)\bigr) =
\begin{cases}
    (0.8,\,0.2,\,0), & t < T_{\mathrm{w}},\\
    (0.4,\,0.4,\,0.2), & t \geq T_{\mathrm{w}}.
\end{cases}
\end{aligned}\end{equation}
During the warm-up stage, the learning rate is linearly increased from zero to its maximum value, and the optimization is dominated by the more noise-stable globally normalized losses. After the model has acquired a preliminary channel representation, a ReduceLROnPlateau scheduler is employed to halve the learning rate with patience 3, and the sample-wise normalized loss $\mathcal{L}_{\mathrm{m3}}$ is introduced to increase the relative contribution of weak channel samples. This stage-wise strategy improves weak channel reconstruction while maintaining robustness to channel-magnitude variations, reducing the risk of excessive sensitivity to low SNR samples.

In addition to the channel reconstruction objectives, we observe that the network may predict multiple weight peaks along the same DoA during optimization, as illustrated in Fig.~\ref{fig_b}. These peaks may reach a spurious equilibrium that degrades prediction accuracy and physical consistency. We therefore introduce a mild regularization on the rendering weights along each ray, where the continuous form depends only on the maximum sampling distance and the rendering weight distribution:
\begin{equation}\begin{aligned}
\label{eq3-20}
\mathcal{L}_{\mathrm{wei1}} (d_f,{\chi}) = \int_0^{d_f} \! \int_0^{d_f} \chi(u) \chi(v) \frac{|u-v|}{d_f} dudv,
\end{aligned}\end{equation}
where $\chi(\cdot)$ denotes the NeRF style volumetric rendering weight density along the ray, which can be interpreted as the continuous function of equivalent weight density $\boldsymbol{\sigma}^{(4)}$. This objective penalizes the pairwise distance between weighted samples and therefore encourages separated or broad peaks to converge toward a compact, approximately unimodal distribution.

It should be noted that $\chi(\cdot)=0$ is a global optimum of $\mathcal{L}_{\mathrm{wei1}}$, which is undesirable for a ray associated with a valid propagation path. Therefore, this regularization is activated slightly only after the main network has entered the stable optimization stage. Let $s_i=d_i/d_f$ denote the normalized boundary of the $i$-th sampling interval. Assuming a uniform weight density within each interval, the discrete objective is
\begin{equation}\begin{aligned}\label{eq3-21}
&\mathcal{L}_{\mathrm{wei2}} (\mathbf{s},\boldsymbol{\chi},t) =\\
&\begin{cases}
0, & t<T_{\mathrm{w}},\\[1mm]
\sum_{u=1}^{m}\sum_{v=1}^{m} \chi_u\chi_v \left|\frac{s_u+s_{u+1}}{2}- \frac{s_v+s_{v+1}}{2}\right|\\
\qquad +\frac{1}{3}\sum_{u=1}^{m}\chi_u^2(s_{u+1}-s_u), & t\geq T_{\mathrm{w}},
\end{cases}
\end{aligned}\end{equation}
Here, $\chi_i$ is the integrated rendering weight of interval $i$. The double sum penalizes separation between intervals, while the second term accounts for within-interval spread. Together they favor compact VA support.

\begin{figure*}[t]
    \centering
    \includegraphics[width=\linewidth]{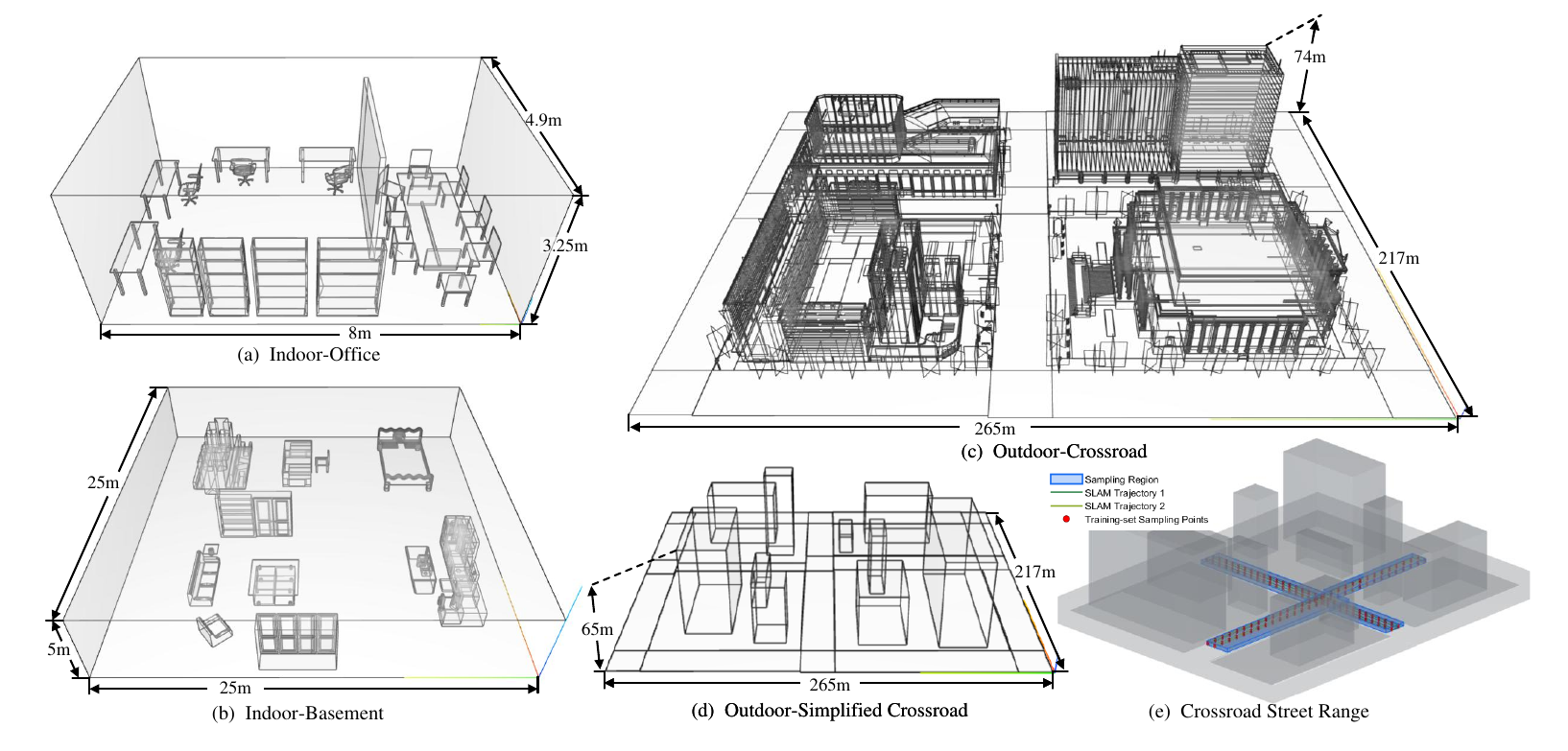}
    \caption{Simulation environments: indoor scenarios consisting of (a) the office and (b) the basement; outdoor scenarios consisting of (c) the crossroad and (d) the simplified crossroad; and (e) the sampling region and predefined receiver trajectory used for multipath-SLAM-assisted prior acquisition.}
    \label{fig_scene}
\end{figure*}

The complete training objective is then given by:
\begin{equation}
\begin{aligned}
\label{eq3-22}
\mathcal{L}_{\mathrm{total}}
\!=\!{}&
\mathcal{L}_{\mathrm{prop}}^{(1)}
\!+\! \mathcal{L}_{\mathrm{prop}}^{(2)}
\!+\! 0.1\mathcal{L}_{\mathrm{m}}^{(3)}
\!+\! 0.9\mathcal{L}_{\mathrm{m}}^{(4)}
\!+\! 0.1\mathcal{L}_{\mathrm{wei2}} .
\end{aligned}
\end{equation}
Here, $\mathcal{L}_{\mathrm{prop}}^{(1)}$ and $\mathcal{L}_{\mathrm{prop}}^{(2)}$ are calculated using \eqref{eq3-10} for the first and second proposal histograms, respectively. In particular, the former is constructed from $(\boldsymbol{\Gamma}^{(1)},\boldsymbol{\sigma}^{(1)})$. Moreover, $\mathcal{L}_{\mathrm{m}}^{(3)}$ and $\mathcal{L}_{\mathrm{m}}^{(4)}$ denote the channel reconstruction losses obtained after the two main network queries. Gradients are stopped during importance resampling from the proposal network to the main network, consequently, the channel reconstruction and weight regularization terms do not update the proposal network, and the two groups of network parameters are optimized independently.

\section{Simulation and Analysis}

\begin{table*}[t]
\centering
\caption{Configurations of the single- and multi-frequency ray-tracing datasets.}
\label{tab:scene_paras}

\resizebox{\textwidth}{!}{%
\begin{tabular}{l c c c c l l}
\toprule

\multicolumn{1}{c}{\multirow{2}{*}{Scenario}}
& \multicolumn{2}{c}{Rx Locations}
& \multicolumn{2}{c}{CFR Samples}
& \multicolumn{2}{c}{Location Sampling Strategy}
\\

\cmidrule(lr){2-3}
\cmidrule(lr){4-5}
\cmidrule(lr){6-7}

& Single-freq.
& Multi-freq.
& Single-freq.
& Multi-freq.
& \multicolumn{1}{c}{Training Set}
& \multicolumn{1}{c}{Validation / Test Set} \\
\midrule

Office (Indoor)
& 3,000
& 1,000
& $3{,}000\times1$
& $1{,}000\times10$
& Random
& Random \\

Basement (Indoor)
& 10,000
& 2,500
& $10{,}000\times1$
& $2{,}500\times10$
& Random
& Random \\

Crossroad (Outdoor)
& 20,000
& 5,000
& $20{,}000\times1$
& $5{,}000\times10$
& Random in designated region
& Random in designated region \\

Simplified crossroad (Outdoor)
& 20,000
& 5,000
& $20{,}000\times1$
& $5{,}000\times10$
& Predefined trajectory (Fig.~\ref{fig_scene}(e))
& Random in designated region \\

\bottomrule
\end{tabular}%
}
\end{table*}

\begin{table*}[t]
\centering
\caption{Main scene-dependent configurations of Prop-WRF.}
\label{tab:network_config}
\resizebox{\textwidth}{!}{%
\begin{tabular}{l c c c c c c c c c}
\toprule
Scenario
& Priors
& PE/IPE Orders
& $d_f$ (m)
& Prop./Main Samples
& Batch Size
& Main LR
& Prop. LR
& Grad. Clip
& Warm-up/Max. Iters \\
\midrule

Office
& GT + DoA noise
& $(10,10,8)$
& 15
& $128/128$
& 128 (64)
& $7\!\times\!10^{-4}$
& $2\!\times\!10^{-4}$
& $5\!\times\!10^{-4}$
& $500/10{,}000$ \\

Basement
& GT + DoA noise
& $(13,13,11)$
& 45
& $256/256$
& 128
& $7\!\times\!10^{-4}$
& $2\!\times\!10^{-4}$
& $2.5\!\times\!10^{-4}$
& $500/10{,}000$ \\

Crossroad
& GT + DoA noise
& $(14,14,12)$
& 160
& $384/384$
& 256 (64)
& $4\!\times\!10^{-4}$
& $2\!\times\!10^{-4}$
& $2.5\!\times\!10^{-4}$
& $800/15{,}000$ \\

Simplified crossroad
& mpSLAM assisted
& $(14,14,12)$
& 180
& $384/384$
& 256
& $4\!\times\!10^{-4}$
& $2\!\times\!10^{-4}$
& $2.5\!\times\!10^{-4}$
& $800/10{,}000$ \\

\bottomrule
\end{tabular}%
}

\vspace{1mm}
\parbox{\textwidth}{\footnotesize \emph{Note:} Multi-frequency experiments are conducted only in the office and crossroad scenarios, for which the parenthesized values denote the corresponding batch sizes. GT and mpSLAM indicate ground truth and multipath based SLAM for short, respectively.}
\end{table*}

\subsection{Datasets Creation}
\subsubsection{Ray-Tracing Scenarios and Datasets}

The datasets are generated using the shooting-and-bouncing-rays (SBR) method implemented in MATLAB R2024a. For each receiver location, the CFR of a single receive antenna is collected at ten subcarrier frequencies ranging from 2.400 to 2.409~GHz with a spacing of 1~MHz. The receiver locations and geometric propagation information are shared across these subcarriers, while the complex CFR at each frequency is separately recorded as the supervision target. All indoor surfaces are modeled as concrete, with relative permittivity $\epsilon_{\mathrm r} \!=\!5$ and conductivity $\sigma_{\mathrm c} \!=\! 0.02~\mathrm{S/m}$, which determine the complex relative permittivity $\eta(f)$. Only specular propagation paths involving at most three reflections are retained.

Fig.~\ref{fig_scene} and Table~\ref{tab:scene_paras} describe the four environments. Office and basement provide indoor benchmarks; crossroad tests outdoor scalability. Simplified crossroad retains a large spatial extent with reduced complexity for controlled evaluation of SLAM-assisted prior acquisition.

\subsubsection{Prior Acquisition}

The geometric inputs to Prop-WRF consist of the query receiver position and its associated candidate DoAs. For the training set, these geometric priors are obtained from the RT-resolved propagation paths. To account for practical parameter-estimation uncertainty, each ground-truth DoA is perturbed by a random angular error whose magnitude is uniformly sampled from $0^\circ$ to $0.1^\circ$. Since the propagation geometry is shared across the considered subcarriers, the same geometric priors are used for all frequency samples at a given receiver location.

For the validation and test sets, the receiver positions are assumed to be known, while the DoA priors are generated following the VA-library strategy adopted in \cite{mip-newrf}. Specifically, a VA library is first constructed from the training set. For each validation or test location, the VAs associated with its nearest training location are retrieved, and the candidate DoAs are obtained from the directions connecting the query receiver to these VAs. This procedure avoids directly using the RT-resolved DoAs at unobserved locations.

\subsubsection{Multipath-SLAM-Assisted Prior Acquisition}\label{slamconfig}

The feasibility of acquiring Prop-WRF training priors through multipath SLAM is evaluated only in the simplified crossroad environment. As shown in Fig.~\ref{fig_scene}(e), the receiver moves along a predefined trajectory with a spacing of approximately 0.128~m between adjacent sampling points. An external positioning device is assumed to provide noisy position observations with a mean Euclidean error of 0.0256~m; nevertheless, the receiver position remains an unknown state to be refined by the SLAM algorithm.

For the multipath-parameter-estimation front end, the receiver is simulated with a $4\times4$ half-wavelength uniform planar array. The array observations contain 80 subcarriers spanning a bandwidth of 400~MHz, with an SNR of 20~dB. Five channel snapshots are collected at each receiver location and jointly processed to estimate the DoAs and ToAs of the dominant multipath components. Before SLAM refinement, the mean DoA error and ToA-equivalent range error are approximately $0.230^\circ$ and 0.017~m, respectively. After exploiting temporal and geometric consistency through multipath SLAM, the mean receiver-position error is reduced to 0.0136~m, while the mean DoA and range errors are reduced to approximately $0.075^\circ$ and 0.011~m, respectively. The RT ground truth is used only to generate the simulated observations and evaluate the estimation errors; it is not provided as an input to the SLAM algorithm.

The resulting training-set construction pipeline consists of four successive stages: array-CFR acquisition, DoA/ToA extraction, multipath-SLAM refinement and VA association, and WRF-prior construction. Let $\widehat{\mathbf{x}}_t$ denote the estimated receiver position and $\widehat{\mathbf{p}}_{k,t}$ denote the estimated position of the $k$-th associated VA at time $t$. The corresponding DoA unit vector and VA-depth prior are obtained as

\begin{equation}
\begin{aligned}
\widehat{\mathbf{d}}_{k,t}
&=
\frac{\widehat{\mathbf{p}}_{k,t}-\widehat{\mathbf{x}}_t}
{\left\|\widehat{\mathbf{p}}_{k,t}-\widehat{\mathbf{x}}_t\right\|_2},\\
\widehat r_{k,t}
&=
\left\|\widehat{\mathbf{p}}_{k,t}-\widehat{\mathbf{x}}_t\right\|_2,
\end{aligned}
\label{eq:slam_derived_priors}
\end{equation}
where $\widehat{\mathbf{d}}_{k,t}$ is used as the DoA input to Prop-WRF and $\widehat r_{k,t}$, which corresponds to the estimated propagation distance, provides a soft depth prior.

When a VA-depth prior is available, a two-level sampling strategy is adopted. Specifically, 50\% of the initial samples are drawn uniformly from the local interval
$[\widehat r_{k,t}-1~\mathrm{m},\,\widehat r_{k,t}+1~\mathrm{m}]$, after truncation to the valid sampling range, while the remaining samples retain the original global sampling range. In this manner, the SLAM-derived depth is not used as exact supervision, preserving the ability of Prop-WRF to recover from imperfect VA estimates. For the validation and test sets, the receiver positions are assumed to be known, and candidate VAs are obtained using the same training-set VA-library association strategy described above.

\begin{figure*}[t]
    \centering
    \includegraphics[width=\linewidth]{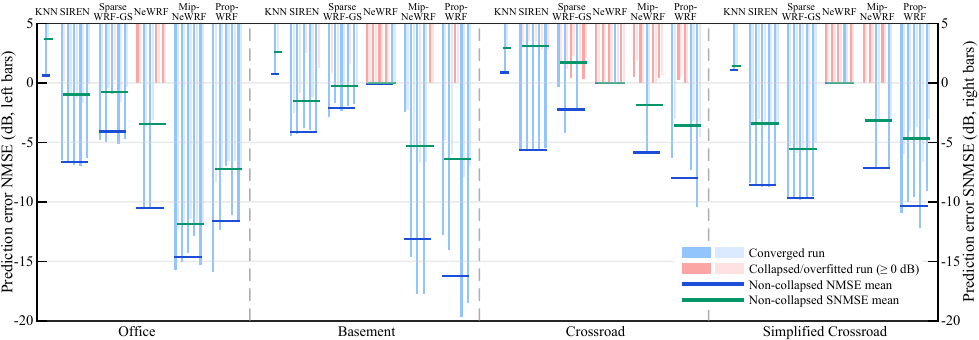}
    \caption{Single-frequency prediction performance of Prop-WRF and five baselines in four scenarios over five random seeds, except for deterministic KNN. Only the simplified crossroad training dataset includes ToA-derived depth priors for initial VA importance sampling.}
    \label{fig_singlefre}
\end{figure*}

\begin{figure*}[t]
    \centering
    \includegraphics[width=\linewidth]{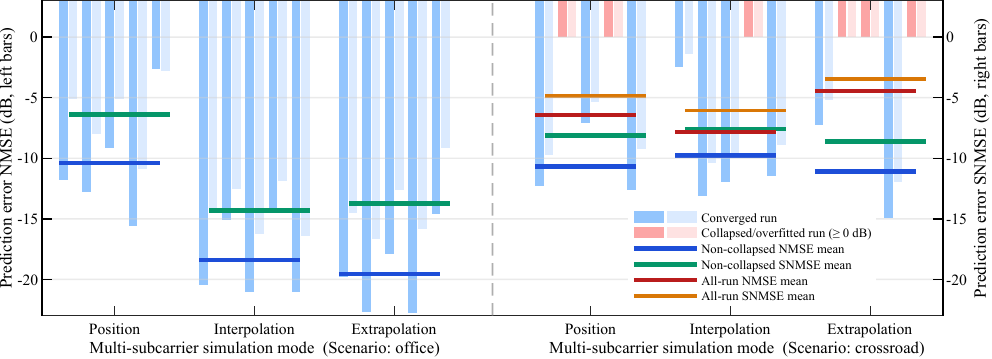}
    \caption{Multi-frequency prediction performance over five random seeds in the office (left) and crossroad (right) scenarios. Horizontal lines denote the mean NMSE and SNMSE; for crossroad, the means over non-collapsed runs are additionally reported.}
    \label{fig_multifre}
\end{figure*}
\subsection{Simulation Setup and Baselines}
For the position-based single/multi-frequency experiments, the receiver locations are randomly divided into training, validation, and test sets with ratios of $70\%$, $10\%$, and $20\%$, respectively. All frequencies at a receiver belong to the same subset to prevent location leakage. For unseen-frequency generalization, the datasets are instead divided along the frequency dimension: $7$, $1$, $2$ of the $10$ subcarriers are used for training, validation, and testing with all available receiver locations retained. Specifically, indices $0$--$9$ correspond to $2.400$--$2.409$~GHz, the interpolation experiment uses indices $\{5\}$, $\{2,8\}$ for validation, testing, and the rest for training, whereas the extrapolation experiment uses $\{8\}$ and $\{0,9\}$ for validation and testing.

Unless otherwise specified, the network structure follows Fig.~\ref{fig_net_arch}. Two proposal-network queries and two main-network queries are performed along each input ray. The directional PE order is set to five, and the cone-radius parameter is $\dot r=0.0017$. For multi-frequency prediction, the frequency PE uses $L_f \!=\! 4$ and is injected only into the complex-radiance branch. The networks are optimized using Adam without weight decay. Linear learning-rate warm-up and gradient clipping are employed to stabilize training. The principal scene-dependent configurations are summarized in Table~\ref{tab:network_config}.

The single-frequency comparison considers five representative baselines. \emph{KNN} estimates the complex CFR by distance-weighted averaging of the CFRs measured at the nearest $k\!=\!5$ training locations. \emph{Phase-referenced SIREN} directly learns the mapping from the receiver position to the complex CFR using MLP with sinusoidal activations~\cite{sitzmann2020siren}. \emph{Sparse WRF-GS} is an input-matched adaptation of WRF-GS~\cite{wrf-gs}, in which Gaussian primitives are initialized from training-set VA candidates and queried only along the supplied dominant propagation directions. \emph{NeWRF} is the original sparse-directional WRF that performs coarse-to-fine neural synthesis along the input DoAs~\cite{newrf}. \emph{Mip-NeWRF} further incorporates scale-consistent hybrid PE--IPE encoding and improved physics-aware channel synthesis~\cite{mip-newrf}. NeRF$^2$ and other unmodified Gaussian-splatting WRFs are not included because they reconstruct dense omnidirectional spatial spectra and require a substantially different input and supervision representation~\cite{nerf2,wrf-gs}. Directly applying them to the present sparse-directional observations would require introducing numerous artificial zero-valued directions and would therefore not constitute a fair comparison. Prediction accuracy is evaluated using the NMSE in \eqref{eq3-15} and the SNMSE in \eqref{eq3-17}, both reported in decibels as $10\log_{10}(\cdot)$. All network related models are trained on a single NVIDIA RTX 4090 GPU, and each experiment is repeated using five random seeds, from 1 to 5, with the mean result reported.

\subsection{Single-Frequency Performance}

Single-frequency training is conducted at 2.400~GHz, with results shown in Fig.~\ref{fig_singlefre}. Prop-WRF generally offers a more favorable trade-off between accuracy and training stability in larger scenarios, although it does not dominate every metric. In office, its NMSE and SNMSE are respectively 3.05 and 4.64~dB higher than those of Mip-NeWRF. A possible explanation is that conventional two stage sampling already adequately covers valid VAs along short rays, whereas additional proposal-guided resampling may introduce distribution approximation errors or miss relevant intervals. As the search range increases, concentrating a limited sampling budget around valid VAs becomes more beneficial. Compared with Mip-NeWRF, Prop-WRF reduces mean errors and yields fewer runs with nonnegative NMSE in basement and two outdoor scenarios, although training instability persists in crossroad. SIREN exhibits relatively consistent NMSE across seeds, but its SNMSE exceeds 0~dB in every crossroad run, indicating that its global NMSE masks poor relative accuracy at weak-channel locations. Sparse WRF-GS performs consistently and achieves the lowest SNMSE in simplified crossroad, where depth priors are available, consistent with their potential benefit for Gaussian initialization.

\subsection{Multi-Frequency Performance}

\begin{figure*}[t]
    \centering
    \includegraphics[width=\linewidth]{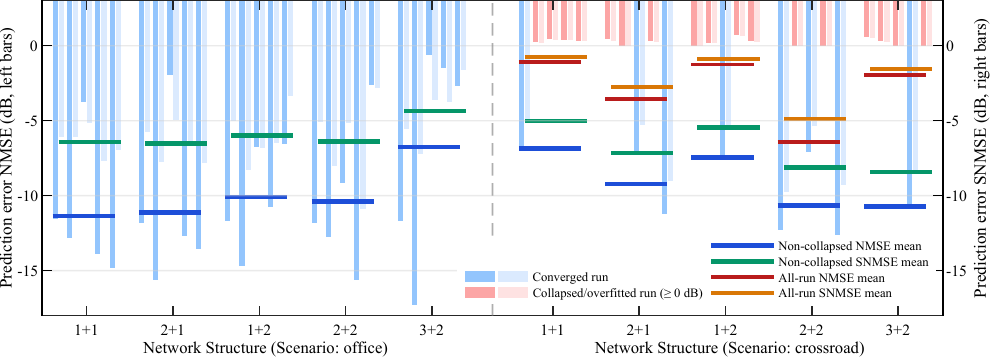}
    \caption{Network structure ablation results over five random seeds in the office and crossroad scenarios.}
    \label{fig_abla_stru}
\end{figure*}
\begin{figure}[t]
    \centering
    \includegraphics[width=\linewidth]{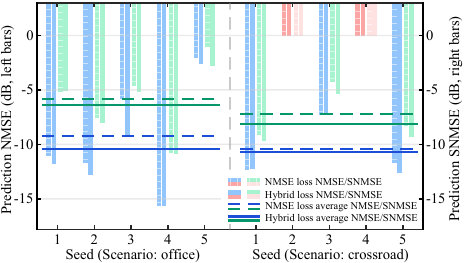}
    \caption{Loss function ablation results in the office and crossroad scenarios.}
    \label{fig_abla_loss}
\end{figure}

Fig.~\ref{fig_multifre} compares spatial generalization, frequency interpolation, and frequency extrapolation of Prop-WRF. In the office scenario, both frequency interpolation and extrapolation clearly outperform spatial generalization, while neither frequency setting consistently dominates under both metrics. This is mainly because the VA geometry remains approximately invariant over the considered $9$-MHz bandwidth and the frequency-based splits retain spatial coverage, allowing the model to reuse the learned geometry and capture only the frequency-dependent complex radiance. In the larger Crossroad scenario, the converged runs exhibit comparable performance across the three settings, but the frequent training collapse prevents a reliable ranking. These results support the proposed separation of frequency-shared VA geometry and frequency-conditioned radiance. They also indicate that optimization stability remains the primary limitation in complex large-scale environments. This is because valid VAs occupy only a small fraction of the long candidate rays, while the channel labels supervise only the coherent sum of all multipath components. Consequently, early sampling errors may weaken the gradients associated with the true VAs and create a self-reinforcing tendency toward the zero-output solution.

\subsection{Ablation Studies}
\subsubsection{Network Structure Ablation}

\begin{figure}[t]
    \centering
    \includegraphics[width=\linewidth]{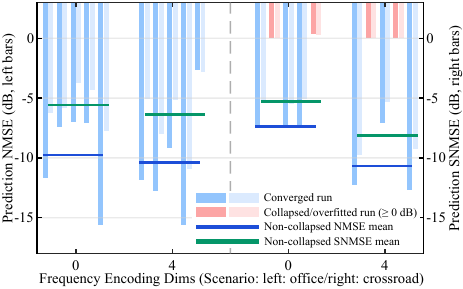}
    \caption{Frequency encoding ablation results in the office and crossroad scenarios.}
    \label{fig_abla_fre_enco}
\end{figure}

As shown in Fig.~\ref{fig_abla_stru}, the benefit of additional proposal/main queries depends strongly on the scene scale. In the office scenario, short propagation ranges and relatively simple VA distributions can already be captured by compact configurations. Additional queries may lead to over-optimization or memorization of the training samples, thereby degrading generalization; nevertheless, when properly optimized, the resulting model can achieve better prediction accuracy. In contrast, the long rays and sparse, complex VA distributions in crossroad require repeated proposal refinement to concentrate samples on effective propagation components. The adopted \(2\!+\!2\) configuration achieves the best stability, with only two collapsed runs compared with the alternatives, while retaining nearly the best converged NMSE and SNMSE. Although \(3\!+\!2\) provides marginally lower errors after successful convergence, its four collapsed runs indicate that excessive refinement strengthens erroneous proposal--main network feedback. Therefore, \(2\!+\!2\) offers the best balance between prediction accuracy and optimization robustness. 

The generally higher SNMSE reflects larger relative errors at weak-channel locations. Occasional reversals indicate poor fitting of strong channels, illustrating the trade-off between global and sample-wise normalization.

\subsubsection{Loss Function Ablation}

To evaluate the adopted hybrid weighted MSE loss, we vary only the loss function with corresponding results shown in Fig.~\ref{fig_abla_loss}. Runs with the same random seed tend to exhibit similar convergence behavior under both losses. Overall, the hybrid loss improves both NMSE and SNMSE, although training collapse persists in the large-scale scenario. This improvement can be attributed to the complementary weighting of strong and weak channel samples: the globally normalized term preserves the contribution of high-power samples, while the additional weighted MSE strengthens supervision for weak channels, yielding a more balanced reconstruction across received-power levels.

\subsubsection{Frequency Encoding Ablation}

As shown in Fig.~\ref{fig_abla_fre_enco}, applying low-order frequency PE improves CFR prediction compared with directly feeding the subcarrier frequency into the radiance MLP (i.e., $L_f=0$), particularly in the large-scale crossroad scenario. The richer sinusoidal features facilitate learning nonlinear frequency-dependent complex radiance. However, excessively high PE orders may promote overfitting or complicate optimization, hindering effective learning within the training budget and occasionally degrading prediction performance.

\subsubsection{Multi-frequency Data Input Ablation}
\begin{figure*}[t]
    \centering
    \includegraphics[width=\linewidth]{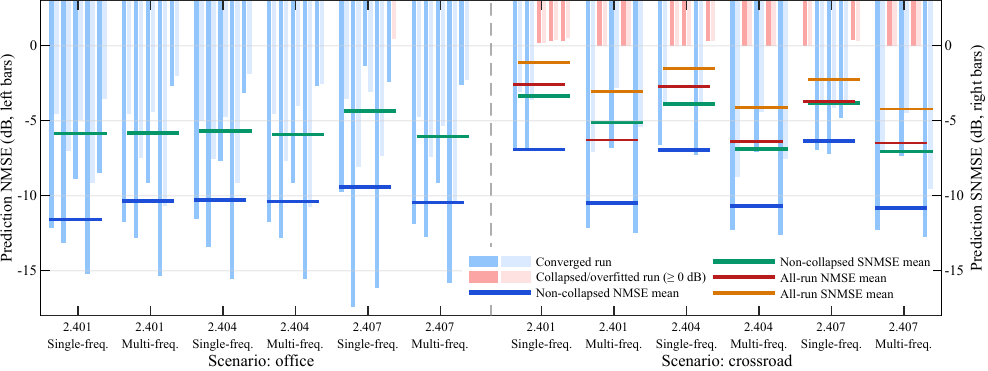}
    \caption{Multi-frequency Data Input ablation results in the office and crossroad scenarios.}
    \label{fig_abla_multifre}
\end{figure*}

Fig.~\ref{fig_abla_multifre} compares single- and multi-frequency training using the same receiver locations. The shared multi-frequency model is trained on all available subcarriers and evaluated at selected frequencies at unseen locations. It generally exhibits more stable performance and lower CFR prediction errors, especially in the large-scale crossroad scenario. This improvement suggests that observations across subcarriers provide complementary constraints on the shared VA geometry, helping mitigate the effects of noisy priors such as DoA errors. In the smaller office scenario, a single-frequency model can already capture the propagation structure effectively, leaving less room for improvement. These results indicate that additional subcarrier observations can improve prediction without increasing the number of sampled receiver locations (although they increase the total amount of CFR training data).

\subsection{Prop-WRF Performance under SLAM-Assisted Dataset}
\begin{figure}[t]
    \centering
    \includegraphics[width=\linewidth]{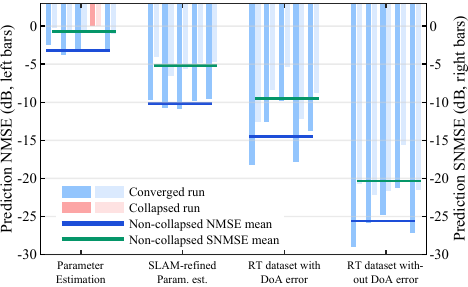}
    \caption{Dataset prior-acquisition comparison over five random seeds in simplified crossroad scenario. All settings use depth priors for importance sampling (see \ref{slamconfig}). The two RT settings assess the gap between idealized RT priors and those obtained through simulated received-signal parameter estimation.}
    \label{fig_slam_prior}
\end{figure}

This experiment evaluates the feasibility of acquiring Prop-WRF training priors instead of establishing a tightly coupled SLAM–WRF system. As shown in Fig.~\ref{fig_slam_prior}, we compare training datasets obtained from received-signal parameter estimation with and without multipath SLAM refinement, alongside RT-based reference datasets to assess the performance gap. With depth priors, relatively consistent performance is exhibited across random seeds, although the persistent gap between NMSE and SNMSE indicates that predicting channels at low-received-power locations remains challenging. Multipath SLAM reduces DoA estimation errors (see subsection~\ref{slamconfig}) and substantially improves prediction accuracy. The remaining gap relative to RT priors with DoA errors may arise from receiver localization errors, VA association and clustering errors, and material misidentification. Removing DoA errors from the RT priors further reduces the average NMSE to $-25.6$~dB, demonstrating accurate DoA estimation as a key limiting factor for sparse-directional WRFs.

\section{Conclusion}
This paper proposed Prop-WRF for multi-subcarrier CFR prediction using proposal-guided VA sampling, frequency-shared geometry, and hybrid normalized losses. Simulations show improved accuracy and stability over conventional sparse-directional WRF baselines in larger scenes, although gains are not uniform and training collapse remains in crossroad. A simulated multipath-SLAM front end further improves prediction by refining geometric training priors. Future work will address training cost, optimization stability, and robustness to DoA errors.




\bibliographystyle{IEEEtran}
\bibliography{ref_}

\vfill

\end{document}